# New hybrid control of a 2 DoF Robot Arm

Mehran Rahmani*, Asif Al Zubayer Swapnil, and Ivan Rulik

*Abstract*— Robot arms have been using in different systems, which the control of designed in desired trajectory is the main task. Also, it is anticipated that while in operation the developed 2DoF robot arm will be constantly encountered with noises such as friction forces. A new integral sliding mode control (NISMC) is therefore being introduced to suppress noise due to its robustness. Then, New hybrid control system (NHISMC) is proposed, which constantly calculates an error value and applies a correction value to the system. This will enhance trajectory and minimize tracking error. In comparison with two other controllers, such as traditional sliding mode control (SMC) and NISMC, experimental results confirmed the efficacy of the proposed control method.

## I. INTRODUCTION

ROBOT manipulator were commonly used in various applications including industrial robots [1] and medical robots [2,3]. One of the most important robot manipulators applications in the area biorobotics is providing robot-aided rehabilitation rehab therapy. For instance, an upper-limb exoskeleton which is a type of robot manipulators is now widely used in providing rehabilitation therapy and/or motion assist. Rahman et al. [4] proposed a new 2DOF robot manipulator mechanism (exoskeleton type) to rehabilitate the elbow and forearm movements. Zhang et al. [5] proposed a new manipulator structure to assist the performance of daily tasks such as eating for people who suffer from the disease. Control of robot manipulators is highly important due to its accuracy for performing its applied works. The SMC approach is one of the well-known control methods that researchers have used because of its robustness and good trajectory tracking efficiency. Adhikary and Mahanta [6] control a structure that does not have a direct drive joint. In order to obtain finite time stability and suitable trajectory tracking performance, a nonsingular sliding mode surface is used.

A time delay control approach is used to achieve partial model independency of the control method. The control solution should have been compared with some other traditional control methods such as PID controller and SMC to show the efficacy of the proposed system. Ferrara and Incremona [7] suggested an algorithm which is suitable for industrial robot applications due to its robustness. A comparison between the proposed algorithm, PID and original suboptimal method demonstrates the effectiveness of that algorithm. However, the control input of each joint did not include in that work to see what will happen when a new control algorithm is applied to the robot. Van et al. [8] provided a SMC for robot manipulator tracking, which offers a finite convergence of time, high robustness and fast transient response. Yao et al. [9] introduced a robust control system to solve the problem of velocity tracking control and position of the automatic train operation (ATO) system. A positive adaptive law is introduced to estimate nonsingular terminal sliding function parameters. To suppress the singularity created by SMC, and make it possible that the velocity tracking error and position tracking error converge to zero, a nonsingular terminal sliding mode control is introduced. In addition, the suggested technique will estimate unknown parameters of the sliding manifold of the ATO framework online. Liu and Yang [10] to guarantee that all the time tracking error converges to a predefined zone, proposed a new error algorithm combined with a novel SMC structure. Furthermore, the controller ensures that steady-state behavior and convergence rate that is more efficient in industrial procedures are based on the input/output data. Simulations for validating theoretical findings are provided.

A new algorithm is applied to estimate the upper bound value to over the limitation of the controller. However, the controller should have applied to a real time system to which results will be obtained. Lee et al. [11] introduced a novel control method and time-delay estimation, which is used to approximate robot dynamics. To eliminate the reaching phase, an integral sliding surface proposed. Then, to obtain high trajectory tracking, an adaptation law is applied. Jin et al. [12] proposed a time delay control algorithm for a robot manipulator. Incremona et al. [13] a robust control approach to address problems with robotic manipulator motion control. To make liable for the matched unmodeled dynamic uncertainties, an integral sliding mode internal loop is proposed. In order to ensures an optimal evolution of the controlled system, an external loop was applied.

Yang et al. [14] applied a decentralized control method for improving tracking of robot manipulator trajectories. Then, to compensate uncertainties, a disturbance observer is proposed. To fix the fast changing components of the uncertainties, an adaptive SMC is introduced. The findings of the simulation were validated the efficacy of the proposed test process.

In this article, a new robot arm control system is developed. To control the proposed robot, a NISMC is used. In addition, a

Mehran Rahmani, Asif Al Zubayer Swapnil, and Ivan Rulik are with the Department of Mechanical Engineering, University of Wisconsin-Milwaukee, WI, CO 53206 USA (email: mrahmani@uwm.edu).

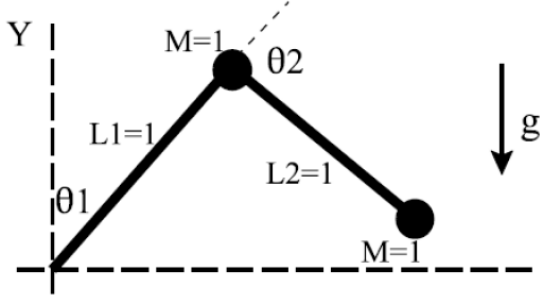

Fig. 1. Schematic robot arm strucutre.

new HNISMC is proposed to continuously estimate error value and reduce it. Experimental tests completely validated the performance of the proposed method.

The majority of the paper is arranged according to: In part 2, the robot structure and generation of the dynamic model is introduced. In part 3, the conventional SMC is included. In part 4, the NISMC is discussed. In part 5, the HNISMC is introduced. Finally, experimental results and discussion are included at the end of the paper.

## II. A 2DOF ROBOT ARM

Fig. 1 illustrates the robot arm. The dynamic modeling of the robot arm is definable as [15]:

$$M(q)\ddot{q} + N(q,\dot{q})\dot{q} + G(q) = \tau \tag{1}$$

where $q, \dot{q}, \ddot{q} \in R^2$ illustrates the position, velocity, and acceleration of the joints, respectively.

Also, in the dynamic model of the 2DOF robot manipulator, $M(q) \in R^{2\times 2}$ represented as the inertia matrix, $N(q,\dot{q}) \in R^{2\times 1}$ known as the vector of centrifugal and Coriolis forces, $G(q) \in R^{2\times 1}$ is a gravitational vector, and $\tau \in R^{2\times 1}$ the joint torques. $M(q), N(q,\dot{q})$, $G(q)$ and are demonstrated in Appendix A.

The Eq. (1) rearranging as:

$$\ddot{q} = M^{-1}(q)(\tau - N(q,\dot{q})\dot{q} - G(q)) \tag{2}$$
$$= M^{-1}(q)\tau - M^{-1}(q)N(q,\dot{q})\dot{q}$$
$$- M^{-1}(q)G(q)$$

The Eq. (2) could be denoted as:

$$\ddot{q} = Hu - J\dot{q} - HG(q) \tag{3}$$

where $H = M^{-1}(q), J = M^{-1}(q)N(q,\dot{q})$, and $u = \tau$. The Eq. (3) therefore becomes:

$$\ddot{q} = (H + \Delta H)u - (J + \Delta J)\dot{q} - (H + \Delta H)G(q) + E(t) \tag{4}$$

where $\Delta H$ and $\Delta J$ represent some uncertainties of parameter variations, $E(t)$ and describes as the external disturbances. The Eq. (4) can be simplified:

$$\ddot{q} = U(t) - J\dot{q} - HG(q) + d(t) \tag{5}$$

where $d(t) = \Delta Hu - \Delta J\dot{q} - \Delta HG(q) + E(t)$ and $U(t) = Hu(t)$

## III. CONVENTIONAL SLIDING MODE CONTROL

SMC is a suitable control system for robustness and tracking performance [16]. Solanes et al. [17] presented an SMC for a human-robot to solve surface treatment purpose including deburring. The major issue in designing the SMC method is how suitably choose sliding mode surface, which can be selected as [18]:

$$\sigma(t) = \dot{e}(t) + \lambda e(t) \tag{6}$$

where $e(t) = q_d - q$ known as the tracking error and $\lambda$ is positive constant. SMC consists of two parts: equivalent control and reaching law control. The equivalent of the conventional SMC can be found according to the coming process.

Taking derivative from Eq. (6) generates:

$$\dot{\sigma}(t) = \ddot{e}(t) + \lambda \dot{e}(t) \tag{7}$$

The Eq. (8) Could be specified by using $\ddot{e}(t) = \ddot{q}_d - \ddot{q}$, and substitute it in Eq.(7) produces

$$\dot{\sigma}(t) = \ddot{q}_d - \ddot{q} + \lambda \dot{e}(t) \tag{8}$$

Substitute Eq. (5) in Eq. (8) generates

$$\dot{\sigma}(t) = \ddot{q}_d - U(t) + J\dot{q} + HG(q) - d(t) + \lambda \dot{e}(t) \tag{9}$$

The equivalent control can be defined by forcing $\dot{\sigma} = 0$.

$$u_{eq}(t) = J\dot{q} + HG(q) - d(t) + \ddot{q}_d + \lambda \dot{e}(t) \tag{10}$$

When external perturbations apply to the system, equivalent control is not able to suppress those noises. A second control named reaching controller is needed in addition to the equivalent control to suppress the external disturbances as:

$$u_r(t) = \Gamma \text{sign}(e(t)) \tag{11}$$

where $\Gamma$ is described as a positive value. Therefore, conventional SMC could be described:

$$u(t) = u_{eq}(t) + u_r(t) \tag{12}$$

The Lyapunov theory denoted as:

$$L(t) = \frac{1}{2}\sigma^T(t)\sigma(t) \tag{13}$$

The stability of the conventional SMC testifies by the Lyapunov theory. The condition which the control system will be stable could be described [19]:

$$\dot{L}(t) = \sigma^T(t)\dot{\sigma}(t) \leq 0, \quad \sigma(t) \neq 0 \tag{14}$$

Substitute Eq. (9) into Eq. (14) generates

$$\dot{L}(t) = \sigma^T(t)(\ddot{q}_d - U(t) + J\dot{q} + HG(q) - d(t) + \lambda \dot{e}(t)) \tag{15}$$

The control input defined by Eq. (12). Substitute it in Eq. (15) produces

$$\dot{L}(t) = \sigma^T(t)(\ddot{q}_d - u_{eq}(t) - u_r(t) \tag{16}$$
$$+ J\dot{q} + HG(q) - d(t) + \lambda \dot{e}(t))$$

Substitute Eq. (10) and Eq. (11) into Eq. (16) generates

$$\dot{L}(t) = \sigma^T(t)(\ddot{q}_d - J\dot{q} - HG(q) + d(t) \tag{17}$$
$$- \ddot{q}_d - \lambda \dot{e}(t) - \Gamma \text{sign}(e(t))$$
$$+ J\dot{q} + HG(q) - d(t) + \lambda \dot{e}(t))$$

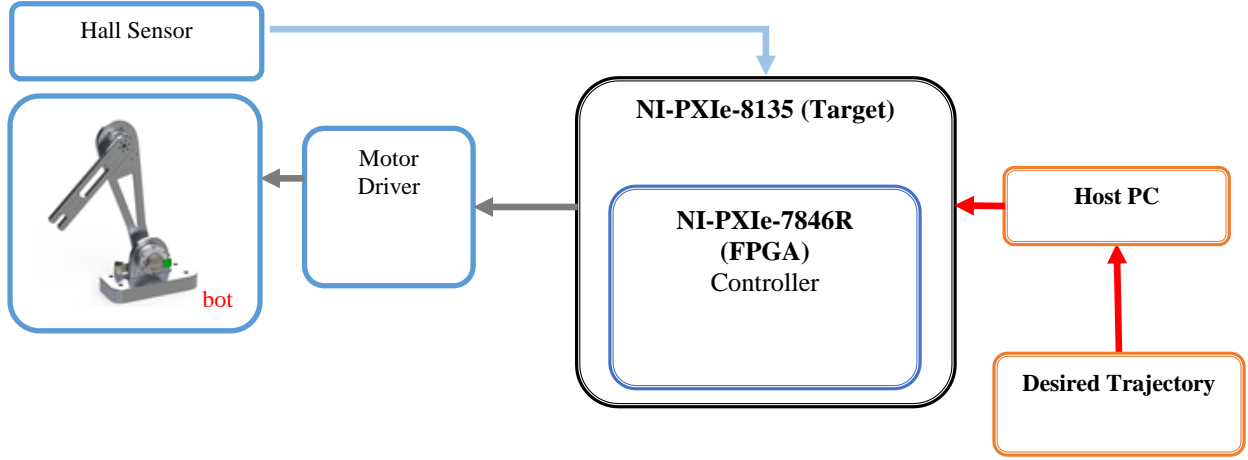

Fig. 2. Experimental results

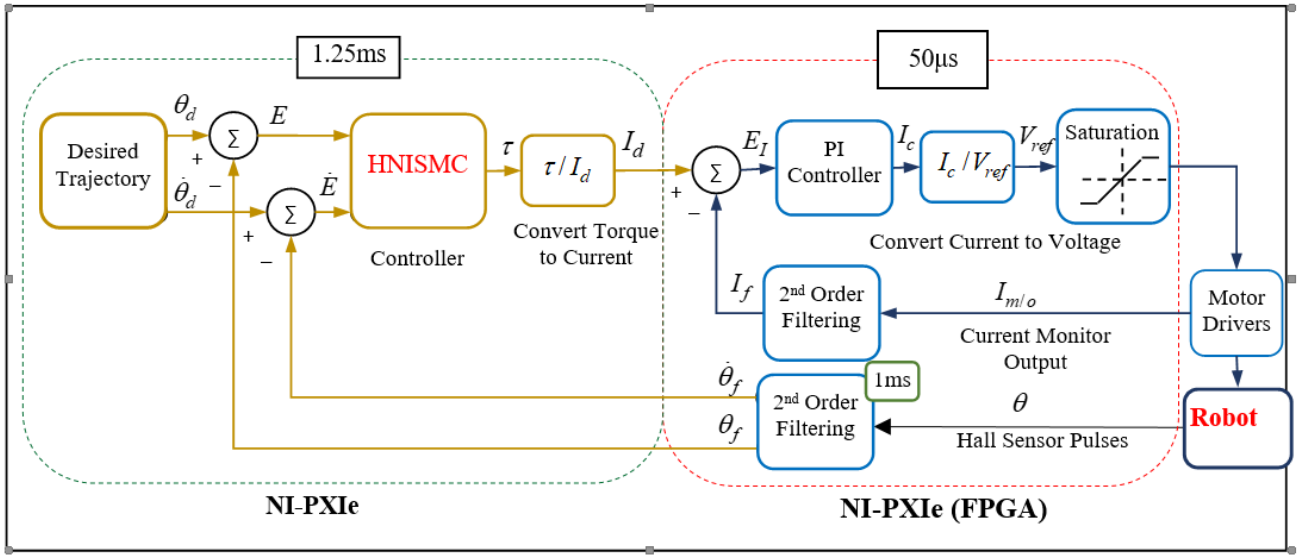

Fig. 3. Control architecture.

Simplify Eq. (17) denotes

$$\dot{L}(t) = \sigma^T(t)(-\Gamma\,\text{sign}(e(t))) \le -\left|\sigma^T(t)\right|\Gamma\,\text{sign}(e(t)) \quad (18)$$

The Eq. (18) satisfied the condition of Eq. (14). Therefore, the controller is stable.

## I. NEW INTEGRAL SLIDING MODE CONTROL

ISMC is better than SMC in terms of stability and low tracking performance [20].

The new integral sliding mode surface could be described as:

$$\sigma(t) = \dot{e}(t) + \int_0^t (\alpha e(\tau) + \beta \dot{e}(\tau))d\tau \quad (19)$$

where $\alpha$ and $\beta$ are positive constants.

The $\dot{\sigma}(t) = 0$ to find equivalent control. Take the Eq. (19) derivative:

$$\dot{\sigma}(t) = \ddot{e}(t) + \alpha e(t) + \beta \dot{e}(t) \quad (20)$$

Substitute $\ddot{e}(t) = \ddot{q} - \ddot{q}_d$ into Eq. (20) produces

$$\dot{\sigma}(t) = \ddot{q}_d - \ddot{q} + \alpha e(t) + \beta \dot{e}(t) \quad (21)$$

Substitute Eq. (5) into Eq. (21) creates

$$\dot{\sigma}(t) = \ddot{q}_d - U(t) + J\dot{q} + HG(q) - d(t) + \alpha e(t) + \beta \dot{e}(t) \quad (22)$$

By using $\dot{\sigma}(t) = 0$, the equivalent controller can be illustrated as:

$$u_{eq}(t) = \ddot{q}_d + J\dot{q} + HG(q) - d(t) \\ + \alpha e(t) + \beta \dot{e}(t) \quad (23)$$

Also, the reaching control law is designed as:

$$u_r(t) = \Gamma\,\text{sign}(e(t)) \quad (24)$$

The control input for NISMC could be created as:

$$u(t) = u_{eq}(t) + u_r(t) \quad (25)$$

By using Eq. (13), Eq. (14), and Eq. (22), the Eq. (26) is:

$$\dot{L}(t) = \sigma^T(t)\dot{\sigma}(t) = \sigma^T(t)(\ddot{q}_d - U(t) + J\dot{q} \\ + HG(q) - d(t) + \alpha e(t) + \beta \dot{e}(t)) \quad (26)$$

Substitute Eq. (25) into Eq. (26) produces

$$\dot{L}(t) = \sigma^T(t)\dot{\sigma}(t) = \sigma^T(t)(\ddot{q}_d - u_{eq}(t) - u_r(t) + J\dot{q} \quad (27)$$
$$+ HG(q) - d(t) + \alpha e(t) + \beta \dot{e}(t))$$

Substitute Eq. (23) into Eq. (27) denotes

$$\dot{L}(t) = \sigma^T(t)\dot{\sigma}(t) = \sigma^T(t)(\ddot{q}_d - \ddot{q}_d - J\dot{q} - HG(q) \quad (28)$$
$$+ d(t) - \alpha e(t) - \beta \dot{e}(t) - u_r(t) + J\dot{q}$$
$$+ HG(q) - d(t) + \alpha e(t) + \beta \dot{e}(t))$$

Simplify Eq. (28) demonstrates

$$\dot{L}(t) = \sigma^T(t)\dot{\sigma}(t) = \sigma^T(t)(-u_r(t)) \quad (29)$$

Substitutes Eq. (24) into Eq. (29) illustrates

$$\dot{L}(t) = \sigma^T(t)(-\Gamma \text{sign}(e(t))) \leq -|\sigma^T(t)|\Gamma \text{sign}(e(t)) \quad (30)$$

The Eq. (30) shows the control approach is stable.

## II. NEW HYBRID INTEGRAL SLIDING MODE CONTROL

Designing a suitable hybrid control is highly effective for the controller. In the present work, a new hybrid controller is proposed. This controller ($u_H(t)$) continuously evaluates the error values and will apply correction values of error. As a result of this process, trajectory tracking will be improved and tracking error will be reduced. The novel controller is described as:

$$u(t) = u_{NISMC}(t) + u_H(t) \quad (31)$$

where the $u_H(t)$ can be shown as:

$$u_H(t) = -\xi_1 \text{sign}(e(t)) - \xi_2 \text{sign}(\dot{e}(t)) \quad (32)$$

where $\xi_1$ and $\xi_2$ are positive constants.

The stability of the proposed HNISMC can be proved by using Eq. (26) as:

$$\dot{L}(t) = \sigma^T(t)\dot{\sigma}(t) = \sigma^T(t)(\ddot{q}_d - U(t) + J\dot{q} \quad (33)$$
$$+ HG(q) - d(t) + \alpha e(t) + \beta \dot{e}(t))$$

Substitute Eq. (31) into Eq. (33) produces

$$\dot{L}(t) = \sigma^T(t)(\ddot{q}_d - u_{NISMC}(t) - u_H(t) + J\dot{q} \quad (34)$$
$$+ HG(q) - d(t) + \alpha e(t) + \beta \dot{e}(t))$$

Substitute Eq. (25) into Eq. (34) denotes

$$\dot{L}(t) = \sigma^T(t)\dot{\sigma}(t) = \sigma^T(t)(\ddot{q}_d - u_{eq}(t) - u_r(t) \quad (35)$$
$$- u_H(t) + J\dot{q} + HG(q) - d(t) + \alpha e(t) + \beta \dot{e}(t))$$

Substitute Eq. (23), Eq. (24), and Eq. (32) into Eq. (35) generates

$$\dot{L}(t) = \sigma^T(t)\dot{\sigma}(t) = \sigma^T(t)(\ddot{q}_d - \ddot{q}_d - J\dot{q} \quad (36)$$
$$- HG(q) + d(t) - \alpha e(t) - \beta \dot{e}(t) - \Gamma \text{sign}(e(t))$$
$$+ \xi_1 \text{sign}(e(t)) + \xi_2 \text{sign}(\dot{e}(t)) + J\dot{q} + HG(q)$$
$$- d(t) + \alpha e(t) + \beta \dot{e}(t))$$

Simplify Eq. (36) demonstrates:

$$\dot{L}(t) = \sigma^T(t)(-\Gamma \text{sign}(e(t)) + \xi_1 \text{sign}(e(t)) \quad (37)$$
$$+ \xi_2 \text{sign}(\dot{e}(t))) \leq |\sigma^T(t)|(-\Gamma \text{sign}(e(t))$$
$$+ \xi_1 \text{sign}(e(t)) + \xi_2 \text{sign}(\dot{e}(t)))$$

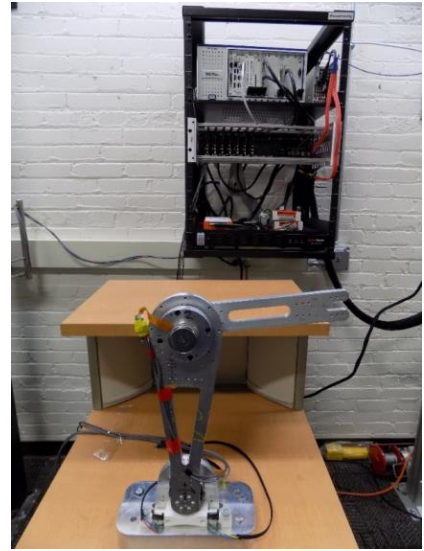

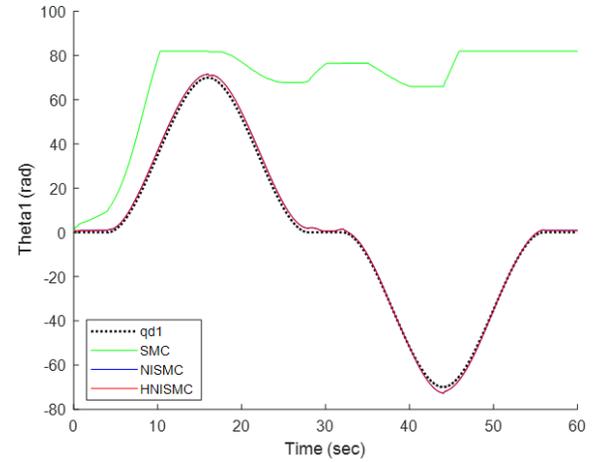

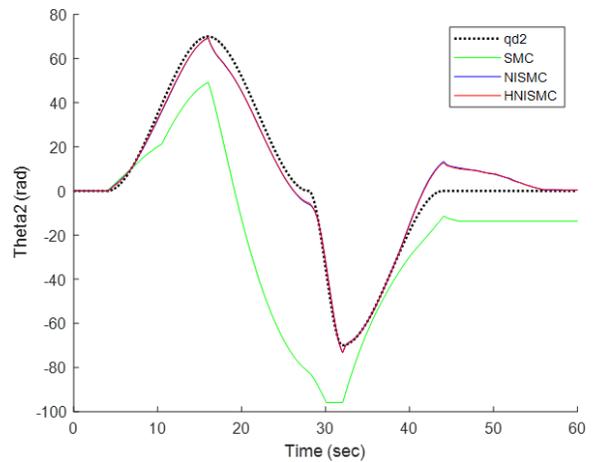

Fig. 5. Position tracking performance of joints under SMC, NISMC, and HNISMC.

The Eq. (37) shows that the proposed controller is stable.

## III. EXPERIMENTAL RESULTS

The experimental schematic structure for the robot arm is shown in Fig. 2. Two maxon motors (EC45) integrated with harmonic drive are attached to the shoulder and elbow joint of the robot to actuate the robot. The motors use built-in Hall sensors to provide the RPM/ angle of rotation. These Hall sensor signals are sampled at 1ms. A second order filter is used to remove high frequency (filtered parameters: $\zeta = 0.9$, and $\omega_0 = 30$ rad/s).

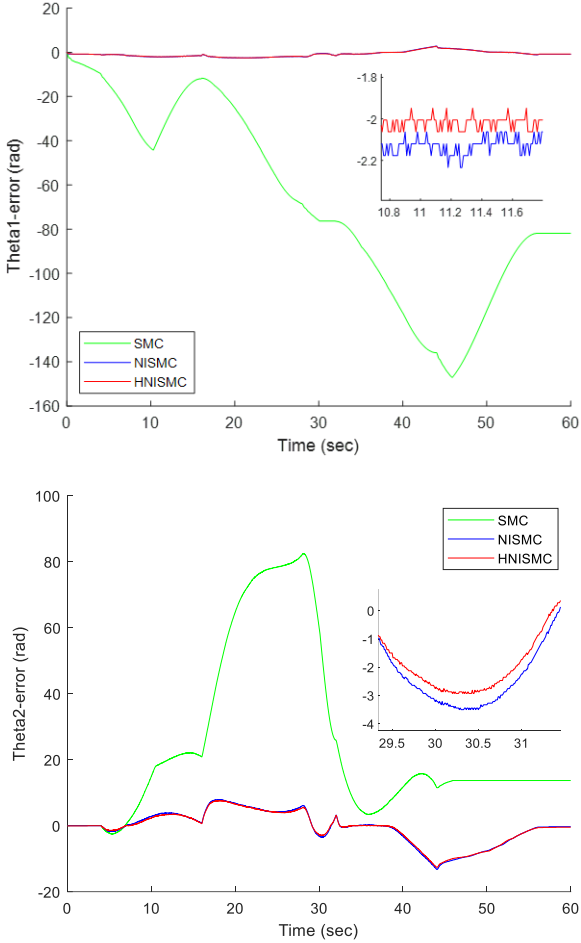

Fig. 6. Position tracking error performance of joints under SMC, NISMC, and HNISMC.

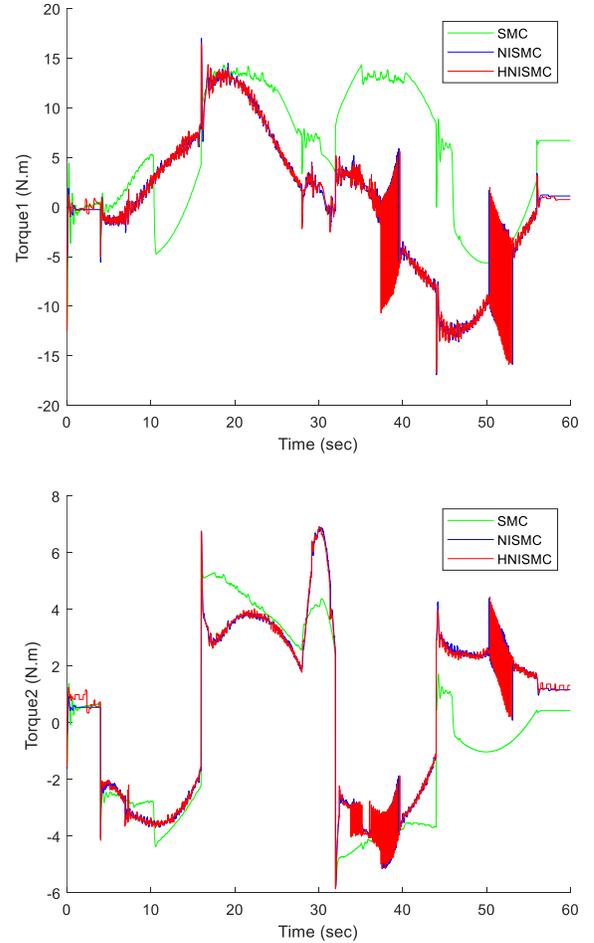

Fig. 7. Control input performance of joints under SMC, NISMC, and HNISMC.

Control architecture for the 2 DOF robot manipulator is depicted in Figure 3, the setup to read the analog signals and apply the desired control through voltage/current pulses its archived by using a host PC running the latest NI LabView® connected via ethernet to a NI (National Instrument) target running a linux-based-real-time OS that allows to process and store information at faster rates and an FPGA module as the main component with a higher sampling rate to read the encoders (Hall Sensors) from the motors and send the control current to the motor drivers. As presented in the controller development section and also in Fig. 3, the controller outputs are the robot joints' torques which are later converted into the current commands for motor drivers. The controller (SMC/NISMC/ HNISMC) runs in NI-PXIe (Figure 3, sampling rate 1.25 ms). As also seen in Fig.3, a low level Proportional controller to control the desired current runs at 50μs (Fig. 3) inside the FPGA. The feedback current signals measured from the motor drivers (at a sampling rate a1ms) are also filtered with a second order filter (sampling parameters: $\zeta$=0.90, and $\omega_0$=3000 rad/s).

prior to being sent to the PI controller. The real time setup is shown in Fig. 4.

In this study, three different controllers are applied in order to control 2DOF robot manipulator suitably. The SMC parameters are selected as $\lambda = \mathrm{diag}\{50,50\}$, and $\Gamma = \mathrm{diag}\{10,10\}$. The NISMC gains are chosen as $\alpha = \mathrm{diag}\{50,50\}$, $\Gamma = \mathrm{diag}\{10,10\}$, and $\lambda = \mathrm{diag}\{800,800\}$. The HNISMC parameters are selected as $\xi_1 = \mathrm{diag}\{0.05,0.05\}$, and $\xi_2 = \mathrm{diag}\{0.05,0.05\}$. The robot structure properties are selected as $L_1$=320mm, $L_2$=360mm, $m_1$=386 gr, and $m_2$=722 gr.

According to Fig. 5, SMC doesn't have good tracking performance. Therefore, a new NISMC is designed to improve trajectory tracking performance. Then, a HNISMC is proposed to improve the effectiveness of the performance control method, which improve trajectory tracking.

Fig. 6 shows the error tracking of the 2 DOF robot manipulator under SMC, NISMC, and HNISMC. SMC control has error tracking, which is not suitable for tracking the desired trajectory. NISMC is much better than SMC, which significantly reduced tracking error. However, a new hybrid controller is proposed that is able to continuously calculated error values and send correction value to the systems. As a result of this, the tracking error will be reduced.

The control input of the joints under SMC, NISMC, and HNISMC are demonstrated in Fig. 7.

IV. CONCLUSION

This paper proposed a new hybrid control system for control a 2DOF robot manipulator. First, a conventional SMC tested experimental on 2 DOF robot manipulator, but it didn't have high tracking performance. Then, a new NISMC is designed to improve trajectory performance and reduce error tracking performance. Also, a new hybrid control method is proposed, which continuously evaluates error value and sends correction value to the system. As a result of this, tracking error reduced. Experimental results verified the performance of the proposed control method in comparison with two other controllers such as SMC and NISMC.

**Appendix A**

$$q = \begin{bmatrix} \theta_1 \\ \theta_2 \end{bmatrix}$$

$$M(q) = \begin{bmatrix} (M_1+M_2)L_1^2 + M_2L_2^2 + 2M_2L_1L_2\cos\theta_2 & M_2L_2^2 + M_2L_1L_2\cos\theta_2 \\ M_2L_2^2 + M_2L_1L_2\cos\theta_2 & M_2L_2^2 \end{bmatrix}$$

$$N(q,\dot{q}) = \begin{bmatrix} -M_2L_1L_2\sin\theta_2(2\dot{\theta}_1\dot{\theta}_2 + \dot{\theta}_2^2) \\ -M_2L_1L_2\sin\theta_2\dot{\theta}_1\dot{\theta}_2 \end{bmatrix}$$

$$G(q) = \begin{bmatrix} -(M_1+M_2)gL_1\sin\theta_1 - M_2gL_2\sin(\theta_1+\theta_2) \\ -M_2gL_2\sin(\theta_1+\theta_2) \end{bmatrix}$$

$$\tau = \begin{bmatrix} \tau_1 \\ \tau_2 \end{bmatrix}$$

V. REFERENCES

[1] Baek, J., Jin, M., & Han, S. (2016). A new adaptive sliding-mode control scheme for application to robot manipulators. IEEE Transactions on industrial electronics, 63(6), 3628-3637.
[2] Seo, H., & Lee, S. (2017, June). Design and experiments of an upper-limb exoskeleton robot. In 2017 14th International Conference on Ubiquitous Robots and Ambient Intelligence (URAI) (pp. 807-808). IEEE.
[3] Kiguchi, K. (2007, June). A study on emg-based human motion prediction for power assist exoskeletons. In 2007 International Symposium on Computational Intelligence in Robotics and Automation (pp. 190-195). IEEE.
[4] Rahman, M. H., Saad, M., Kenné, J. P., & Archambault, P. S. (2010, June). Exoskeleton robot for rehabilitation of elbow and forearm movements. In 18th Mediterranean Conference on Control and Automation, MED'10 (pp. 1567-1572). IEEE.
[5] Zhang, X., Wang, X., Wang, B., Sugi, T., & Nakamura, M. (2008, October). Real-time control strategy for EMG-drive meal assistance robot—My spoon. In 2008 International Conference on Control, Automation and Systems (pp. 800-803). IEEE.
[6] Adhikary, N., & Mahanta, C. (2018). Sliding mode control of position commanded robot manipulators. Control Engineering Practice, 81, 183-198.
[7] Ferrara, A., & Incremona, G. P. (2015). Design of an integral suboptimal second-order sliding mode controller for the robust motion control of robot manipulators. IEEE Transactions on Control Systems Technology, 23(6), 2316-2325.
[8] Van, M., Mavrovouniotis, M., & Ge, S. S. (2018). An adaptive backstepping nonsingular fast terminal sliding mode control for robust fault tolerant control of robot manipulators. IEEE Transactions on Systems, Man, and Cybernetics: Systems, 49(7), 1448-1458.
[9] Yao, X., Park, J. H., Dong, H., Guo, L., & Lin, X. (2018). Robust adaptive nonsingular terminal sliding mode control for automatic train operation. IEEE Transactions on Systems, Man, and Cybernetics: Systems, 49(12), 2406-2415.
[10] Liu, D., & Yang, G. H. (2017). Data-driven adaptive sliding mode control of nonlinear discrete-time systems with prescribed performance. IEEE Transactions on Systems, Man, and Cybernetics: Systems.
[11] Lee, J., Chang, P. H., & Jin, M. (2017). Adaptive integral sliding mode control with time-delay estimation for robot manipulators. IEEE Transactions on Industrial Electronics, 64(8), 6796-6804.
[12] Jin, M., Kang, S. H., Chang, P. H., & Lee, J. (2017). Robust control of robot manipulators using inclusive and enhanced time delay control. IEEE/ASME Transactions on Mechatronics, 22(5), 2141-2152.
[13] Incremona, G. P., Ferrara, A., & Magni, L. (2017). MPC for robot manipulators with integral sliding modes generation. IEEE/ASME Transactions on Mechatronics, 22(3), 1299-1307.
[14] Yang, Z. J., Fukushima, Y., & Qin, P. (2011). Decentralized adaptive robust control of robot manipulators using disturbance observers. IEEE Transactions on Control Systems Technology, 20(5), 1357-1365.
[15] Jin, M., Lee, J., Chang, P. H., & Choi, C. (2009). Practical nonsingular terminal sliding-mode control of robot manipulators for high-accuracy tracking control. IEEE Transactions on Industrial Electronics, 56(9), 3593-3601.
[16] Mancini, M., Bloise, N., Capello, E., & Punta, E. (2019). Sliding mode control techniques and artificial potential field for dynamic collision avoidance in rendezvous maneuvers. IEEE Control Systems Letters, 4(2), 313-318.
[17] Solanes, J. E., Gracia, L., Muñoz-Benavent, P., Miro, J. V., Girbés, V., & Tornero, J. (2018). Human-robot cooperation for robust surface treatment using non-conventional sliding mode control. ISA transactions, 80, 528-541.
[18] Kasera, S., Kumar, A., & Prasad, L. B. (2017, March). Analysis of chattering free improved sliding mode control. In 2017 International Conference on Innovations in Information, Embedded and Communication Systems (ICIIECS) (pp. 1-6). IEEE.
[19] Rahmani, M., Ghanbari, A., & Ettefagh, M. M. (2016). Robust adaptive control of a bio-inspired robot manipulator using bat algorithm. Expert Systems with Applications, 56, 164-176.
[20] Rahmani, M., & Rahman, M. H. (2018). Novel robust control of a 7-DOF exoskeleton robot. PloS one, 13(9), e0203440.